\documentclass[superscriptaddress,showkeys,showpacs,runinaddress,twocolumn,pre]{revtex4}

\usepackage{textcomp,graphicx}  

\newcommand{\ccite}[1]{\cite{#1}}
\newcommand{\scite}[1]{~\cite{#1}}
\renewcommand{\ccite}[1]{\scite{#1}}

\newcommand{\mcite}[1]{\ccite{#1}}

\newcommand{\ecite}{}

\newcommand{\ack}{\section{Acknowledgments}}
\renewcommand{\ack}{\begin{acknowledgments}}

\usepackage{amsmath,amssymb}

\newcommand{\slead}{Section~}  
\newcommand{\flead}{Figure~} 
\newcommand{\sref}[1]{\slead\ref{#1}} 
\newcommand{\fref}[1]{\flead\ref{#1}} 
\newcommand{\sfig}[1]{\textit{(#1)}~}
 
\newcommand{\latin}[1]{\textit{#1}}
 \newcommand{\eg}{\latin{e.g.}, }
 \newcommand{\etal}{\latin{et al.}}

 \newcommand{\mAA}{\text\AA}

 \newcommand{\ee}[2]{{#1}\times10^{#2}}

 \DeclareMathOperator{\erf}{erf}

\newcommand{\avg}[1]{\left\langle{#1}\right\rangle}

\newcommand{\cv}{u_{p,50}}
\newcommand{\rup}{$r$-$u_p$ pair}

\newcommand{\co}{(Color~online)~}

\renewcommand{\latin}[1]{#1}

\newcommand{\mpunct}[1]{}
\renewcommand{\mpunct}[1]{#1}

\newcommand{\eqqref}[1]{Eq.~\eqref{#1}}
\newcommand{\Fref}[1]{Figure~\ref{#1}}
\renewcommand{\flead}{Fig.~}

\begin{document}                

\newlength{\figwidth}
\setlength{\figwidth}{\linewidth}

\title{Critical velocities for deflagration and detonation triggered by voids in a REBO high explosive}

\author{S.~Davis~Herring}
\email{herring@lanl.gov}
\affiliation{Theoretical~Division, Los~Alamos~National~Laboratory, Los~Alamos,~NM~~87545}
\affiliation{Department~of~Applied~Science, University~of~California,~Davis, Davis,~CA~~95616}

\author{Timothy~C.~Germann}
\email{tcg@lanl.gov}
\affiliation{Theoretical~Division, Los~Alamos~National~Laboratory, Los~Alamos,~NM~~87545}

\author{Niels~Gr{\o}nbech-Jensen}
\email{ngjensen@ucdavis.edu}
\affiliation{Department~of~Applied~Science, University~of~California,~Davis, Davis,~CA~~95616}

\pacs{62.50.Ef, 47.40.Nm, 82.40.Fp}
\keywords{shock sensitivity, high explosives, molecular dynamics, REBO}

\begin{abstract}
The effects of circular voids on the shock sensitivity of a two-dimensional model high explosive crystal are considered.  We simulate a piston impact using molecular dynamics simulations with a Reactive Empirical Bond Order (REBO) model potential for a sub-micron, sub-ns exothermic reaction in a diatomic molecular solid.  The probability of initiating chemical reactions is found to rise more suddenly with increasing piston velocity for larger voids that collapse more deterministically.  A void with radius as small as 10 nm reduces the minimum initiating velocity by a factor of 4.
The transition at larger velocities to detonation is studied in a micron-long sample with a single void (and its periodic images).  The reaction yield during the shock traversal increases rapidly with velocity, then becomes a prompt, reliable detonation.  A void of radius 2.5 nm reduces the critical velocity by 10\% from the perfect crystal.  A Pop plot of the time-to-detonation at higher velocities shows a characteristic pressure dependence.
\end{abstract}

\maketitle

\section{Introduction}
Microscopic defects in solid high explosives can have dramatic effects on the sensitivity of the bulk material to initiation by shock waves\ccite{bowden/yoffe}.  When a shock that is insufficiently strong to ignite the material directly encounters a defect, reflection and refraction redirect its energy.  Where the energy density is reduced, little changes; the shock was already inert.  However, a local, temporary increase in energy density may drive exothermic chemical reactions; the resulting hotspot will not generally cause detonation directly, but will increase the pressure behind the shock and thus its strength.  As the shock increases in strength, more energy is available for focusing and less focusing is needed to initiate further reactions.  The positive feedback that results is the principal driver of the transition from shock to detonation in heterogeneous explosives\ccite{beyond-standard}.  The details of the defect feedback process remain poorly understood\mcite{beyond-standard}, and so practical questions like ``To what extent would a population of voids (with some distributions in space and size) reduce the critical velocity of this explosive?''\ go unanswered\ecite.

\begin{figure}
\centering\includegraphics[width=\linewidth]{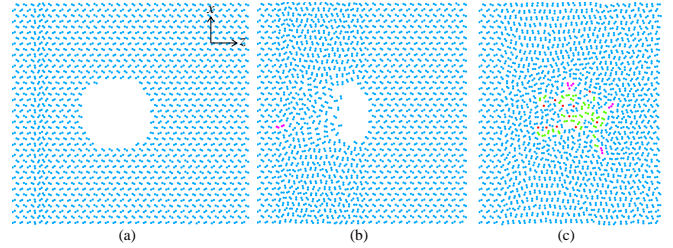}
\caption{\co Snapshots from a $124\times135$~\AA, 2372-atom low-velocity simulation with $r=20$~\AA\ and $u_p\approx3$~km/s.  Undisturbed material at left is the piston; green, red, and purple atoms are products, radicals, and clusters respectively.  \sfig aJust after impact.  \sfig b200~fs before the void finishes collapsing.  \sfig cEnd of the simulation; shock has reached the free surface 1.17~ps after impact.}
\label{cv0seq}
\end{figure}

Spherical voids are a common defect in cast and formed explosives\ccite{zukas/walters}.  When a sufficiently strong shock wave encounters a void, the leading surface is ejected into the void, and the resulting gas is compressed as the void collapses; jetting may also occur\ccite{bowden/yoffe,explosives/propellants,beyond-standard}.  The shock's energy is focused onto the downstream pole of the void.  If the void is large enough and the shock strong enough, chemical reactions result.  This process is illustrated in \fref{cv0seq}.

Several attempts to identify the atomistic mechanism of chemical initiation in collapsing voids have been made using molecular dynamics (MD) simulations\ccite{extreme-dynamics}.  Mintmire \etal\mcite{info:lanl-repo/inspec/4735043}\ determined that energy was efficiently transferred into the molecular vibrational modes of a nonreactive diatomic molecular solid only when the collapse of the void was turbulent and involved the disintegration of its walls; such vibrational excitation is thought to be a necessary precursor to chemical reaction\ecite.  White \etal\mcite{info:lanl-repo/eix/EIX96233135504}\ found that randomly placed circular voids significantly affected the response of ozone crystals under shock loading\ecite.  Germann \etal\mcite{germann13}\ observed that reactions occurred some time after the ejecta collided with the downstream wall (and, with a periodic array of voids, could lead to detonation), and that the reduction in critical shock strength for ignition depended on the orientation as well as the size of elliptical voids\ecite.  In particular, sensitivity was observed to increase with the width of a planar gap, suggesting that heating via recompression of the ejected gas was important for initiation.  Holian \etal\mcite{info:lanl-repo/inspec/7579851
}\ extended the planar gap analysis with a Lennard-Jones potential and derived an expression for the temperature increase due to recompression\ecite.

Hatano\mcite{info:lanl-repo/inspec/7837850} considered the non-equilibrium mechanics of the ejected material in cuboidal voids and observed that the frequency of energetic molecular collisions reached a maximum after temperature did and could vary independently of that maximum\ecite.  Shi and Brenner\mcite{info:lanl-repo/inspec/10238759} considered infinite rectangular voids in N$_8$ cubane crystals and observed reactions following almost immediately after the initial downstream jet impact, with individual molecular collisions at the impact point leading directly to dissociation\ecite.  Turbulent destruction of the void walls and focusing of the ejecta by the walls were observed to increase sensitivity in that system, but recompression after the jet impact (or in the absence of any jet) was not.

\begin{figure*}
\centering\includegraphics[width=\linewidth]{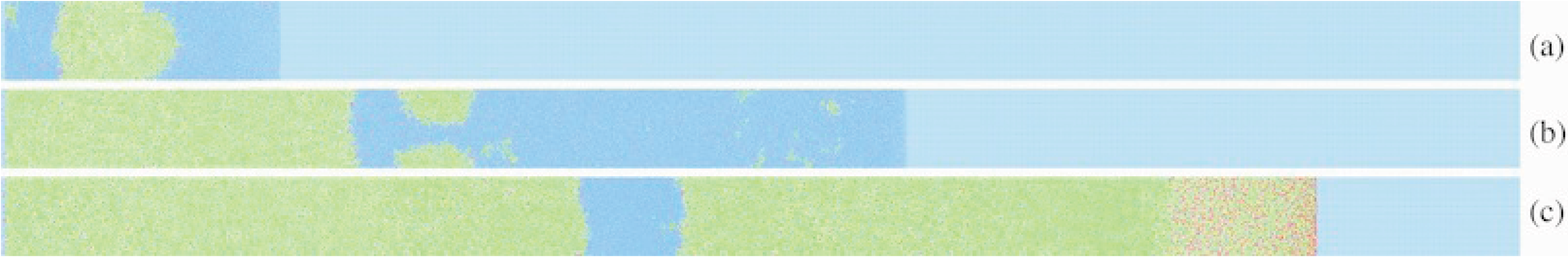}
\caption{\co Snapshots of the forward $\sim$40\% of a $204\times10013$~\AA, 313086-atom high-velocity simulation with $r=50$~\AA\ and $u_p\approx4$~km/s; colors as in \fref{cv0seq}.  \sfig aThe periodic images of the hotspot have merged.  \sfig bReactions develop between the deflagration and the shock front.  \sfig cDetonation has commenced, separated from the original deflagration; the zone of increased dissociation near the shock corresponds to the overdriving in \fref{shockex}.}
\label{hvseq}
\end{figure*}

To better understand the output of void-based hotspots (as a function of void size and input shock strength) and their ability to precipitate detonation, we use MD simulations of supported shocks in an assortment of two-dimensional samples having one circular void (or, more precisely, a periodic line of voids) each.  In this initial exploration we consider only one generation of hotspots, excluding the feedback process between a shock and the sequence of voids it encounters.  For high enough shock strengths, however, a similar feedback can develop among scattered reactions in the shocked material.  A similar transition to detonation thus occurs nonetheless, as depicted in \fref{hvseq}.

\section{Method}
\subsection{Approach}
We perform two separate investigations in different velocity regimes to study the transitions, as radius and/or velocity increase, from no reactions at all to slow deflagration and then to a detonation wave.

Low-strength shocks may collapse a void without producing any chemical reactions.  We simulate low-velocity pistons impacting samples with voids of various radii; \fref{cv0seq} is taken from one such simulation.  In each case, several initial conditions with varied random thermal velocities are considered so as to obtain a probability of initiation.  Very small voids (and the special case of zero void size) are excluded from this study because spontaneous reactions would compete with those triggered by the void and spoil the results.

At higher impact velocities, the hotspot always reacts but may produce a detonation wave only much later if at all.  It is difficult to establish with a finite sample that a detonation would never develop from an observed deflagration, but the progress of the deflagration at moderate velocities and the promptness of the detonation at high velocities may be used to bracket the critical velocity.  The high-velocity piston impact simulations are similar to those in the low-velocity study: they involve various void radii and start from several thermally random initial conditions for each case.  \Fref{hvseq} is taken from one such simulation in which the sample detonated.

The stochastic initiation process is studied in samples having voids of radius 1 to 10~nm with piston velocities of up to 3.5~km/s.  The high-velocity study of the transition to detonation involves samples having voids of radius 1, 2.5, or 5~nm as well as the perfect-crystal case ($r=0$).  There, the pistons have velocities of $u_p=2.95$--4.90~km/s and produce pressures of 10.8--20.9~N/m.  Such two-dimensional pressures are difficult to interpret physically, but following\scite{info:lanl-repo/inspec/4065531} we may suppose that $1\text{~N/m}\approx2.5$~GPa.

\subsection{Model}
The Reactive Empirical Bond Order (REBO) ``AB'' potential (originally developed in\scite{info:lanl-repo/inspec/4065531,info:lanl-repo/inspec/4411199,
extreme-dynamics}) describes an exothermic $\mathrm{2AB \rightarrow A_2 + B_2}$ reaction in a diatomic molecular solid and exhibits typical detonation properties but with a sub-micron, sub-ns reaction zone that is amenable to MD space and time scales.  Heim \etal\ modified it to give a more molecular (and less plasmalike) Chapman-Jouguet state\ccite{info:lanl-repo/isi/000260573900087}.  We utilize the SPaSM (Scalable Parallel Short-range Molecular dynamics) code\mcite{info:lanl-repo/eixxml/1994121436059} and that modified REBO potential (``ModelIV'')\ecite.  A standard leapfrog-Verlet integrator is used with a fixed timestep of 0.509~fs in the NVE ensemble.

Each two-dimensional sample is a rectangle of herringbone crystal with two AB molecules in each $6.19\times4.21$~\AA\ unit cell.  The shock propagates in the $+z$ direction; the samples are periodic in the transverse $x$ direction.  A circular void is created by removing all dimers whose midpoints lie within a circle of a given radius (see \fref{cv0seq}(a)).  In the low-velocity cases, the sample is made large enough to prevent interaction between the periodic images of the void until the collapse is finished and the material has or has not reacted.  Depending on the size of the void, 672--49700 atoms are simulated.  The high-velocity samples are 1~\textmu m (2381 unit cells) long and two void diameters wide (or 10~nm in the case of no void), with the void center four diameters from the piston face; they include 66622--313086 atoms.

Three layers of unit cells at the $-z$ end are frozen to serve as a piston (of infinite mass), and the rest is assigned a tiny but finite temperature (low velocity:~11.6~mK, high velocity:~1.00~mK) and a bulk velocity $v_z=-u_p$ directed into the piston.  (The temperatures must be small to avoid melting the material: the 5~meV depth of the van der Waals well corresponds to just 58~K.)  To reduce the correlation between the different histories, the initial thermal velocities are allowed to thermalize for 1~ps (5~ps for the low-velocity study) before the bulk velocity is applied.

\subsection{Analysis}
We define two atoms as bound if one does not have escape velocity with respect to the other (taking the potential's repulsive core into account\ccite{info:lanl-repo/isi/000260573900087}).  At regular intervals, the atoms to which each atom is bound are noted.  Two atoms each bound only to the other are deemed a molecule; heteronuclear AB molecules are reactants and homonuclear molecules are products.  Many of the results derive from a count of such reacted molecules.  Atoms bound but not in a molecule are termed clusters; \fref{cv0seq}(c) involves all four possible labels.

To identify in which simulations and at which times detonations develop, we measure the position of the shock wave (whether reacting or not) at intervals of 51~fs throughout each simulation.  The shock positions are obtained by finding the pair of adjacent columns of well-populated computational cells (of thickness $\Delta z\approx0.53$~nm) with the largest difference in $\avg{v_z}$.  The identified positions are thus only accurate to $\Delta z$ and are occasionally much too small (when some local fluctuation in the shocked region is misidentified as the shock).  Before seeking the detonation transition, the positions are filtered by removing all values smaller than any preceding them.

Detonation transition times are extracted from the filtered shock positions by finding $\bigl((t_1,z_1),(t_2,z_2),(t_3,z_3)\bigr)$ triples with $z_3-z_2$ and $z_2-z_1$ each $\gtrsim10$~nm that maximize the weighted second derivative
\begin{equation}
\sqrt{m_2}\frac{m_2-m_1}{t_3-t_1}\qquad\left(\mpunct{\text{with }}m_i:=\frac{z_{i+1}-z_i}{t_{i+1}-t_i}\right)\mpunct,
\end{equation}
where the additional $\sqrt{m_2}$ slope factor favors the transition to detonation over any sudden acceleration associated with mere deflagration.  Maximum scores greater than a manually chosen threshold of $\ee{3.9}{12}\text{~m}^{3/2}\text{s}^{-5/2}$ are taken to indicate detonation transitions.  (At low velocities, the void collapse can be mistaken for a transition.  These false positives are easily identified by the large gap between them and the true detonations.)

\begin{figure}
\centering\includegraphics[width=\figwidth]{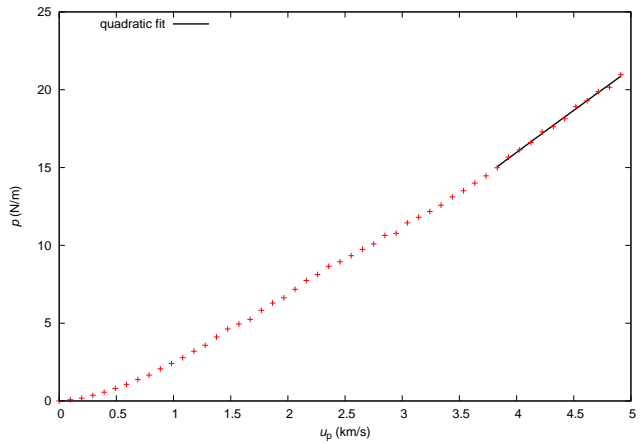}
\caption{Hugoniot for the unreacted material for piston velocities through the largest used in the high-velocity study.  The line is a smoothing quadratic fit through the data points in its interval.}
\label{hug}
\end{figure}

We also determine the shock pressure $p$ associated with a piston velocity $u_p$ from the shock position data.  We measure the shock velocities $u_s$ from the beginnings of voidless simulations (to minimize the effects of reactions on the shock positions) and calculate $p=\rho_0u_su_p$ (the Hugoniot jump condition with $p_0=0$).  For reference, the resulting unreacted Hugoniot is shown in \fref{hug}; the pressures used in \sref{hv} are smoothed by sampling from a quadratic fit to the highest-velocity data.  (The product Hugoniot is presented in\scite{info:lanl-repo/isi/000260573900087}.)

\section{Results}
\subsection{Low-velocity regime: probability of initiation}
\begin{figure}
\centering\includegraphics[width=\figwidth]{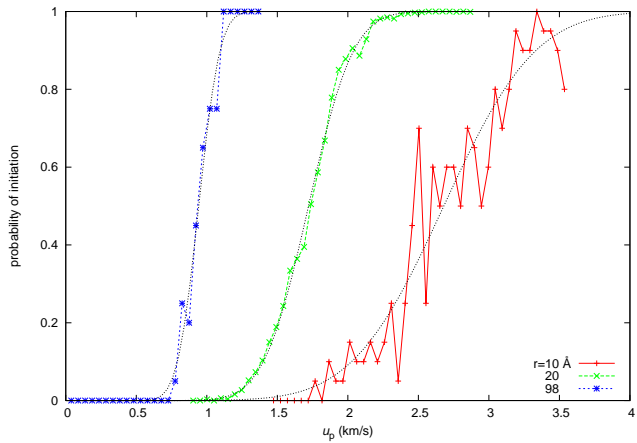}
\caption{Probabilities of initiation as a function of piston velocity at three selected void radii, each with a sigmoid fit.  Larger voids shift the transition to lower velocities and make it sharper.  At $r=20$~\AA, 1000 simulations were performed per velocity, so the statistical noise is much reduced.}
\label{probex}
\end{figure}

Initiation probabilities were derived from at least 20 realizations of each of 1560 radius-velocity pairs (73553 simulations total).  The probabilities obtained for three radii (the largest, the smallest, and an intermediate value with high-quality data) are given in \fref{probex}.
At each radius (10, 12, 14, \ldots, 98~\AA), the critical velocity~$\cv$ (as explored in\scite{germann13}) and transition sharpness~$a$ were determined by scaling and shifting the sigmoid function $P_0(x):=(1+\erf x)/2$ in velocity to fit the measured probabilities:
\begin{equation}
P(u_p)\approx\frac{1+\erf a(u_p-\cv)}2\mpunct.
\label{sssig}
\end{equation}
(Other similar functions [\eg the logistic function $L(x):=(1+e^{-x})^{-1}$] were considered; $P_0(x)$ was selected because its width-slope product was judged most similar to that in the data.)  \Fref{probex} also contains the sigmoid fits for its three radii.

\begin{figure}
\centering\includegraphics[width=\figwidth]{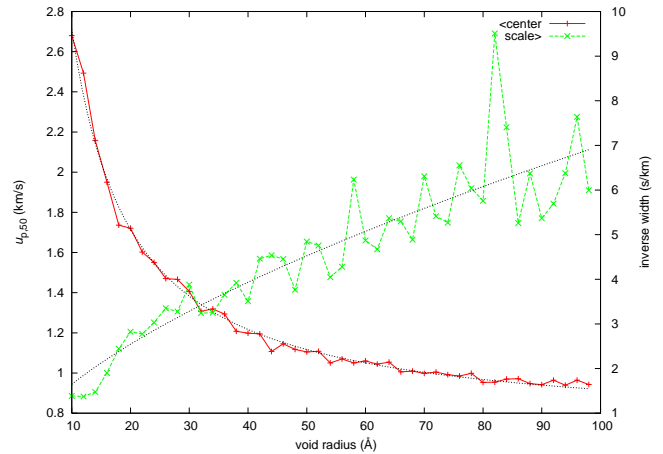}
\caption{Fit parameters [see~\eqqref{sssig}] for the initiation probability as a function of piston velocity at each void radius, with fits to them described in the text.  Larger inverse widths correspond to more sudden transitions from no reactions to guaranteed reactions.}
\label{sigparam}
\end{figure}

The parameters in \eqqref{sssig} for all radii are given in \fref{sigparam}.  Also shown are fits to the center (the 50\% contour) and scale parameters as functions of radius:
\begin{align}
\cv(r)&\approx\frac\alpha r+\beta \label{cfit}\\
\mpunct{\intertext{and}}
a(r)&\approx\gamma\left(\frac r\mAA\right)^\delta\mpunct,
\end{align}
where $\alpha=\ee{1.99}{-6}$~m$^2$/s, $\beta=718$~m/s is the asymptotic value for large voids, $\gamma=\ee{3.88}{-4}$~s/m, and $\delta=0.628$.  The critical velocity for initiation decreases with radius, as the void focuses more of the shock and creates higher temperatures\ccite{germann13,info:lanl-repo/inspec/7579851
}; at $r=10$~nm it is a factor of 4 lower than the minimum velocity observed to initiate reactions in a voidless sample.  (Several other models were considered for the critical velocity, but \eqqref{cfit} was far more successful than the other fits.)  The width (in velocity) of the transition also decreases with increasing void radius, as the hotspot development becomes less dependent on the stochastic behavior of individual molecules.

\subsection{High-velocity regime: transition to detonation}
\label{hv}
\begin{figure}
\centering\includegraphics[width=\figwidth]{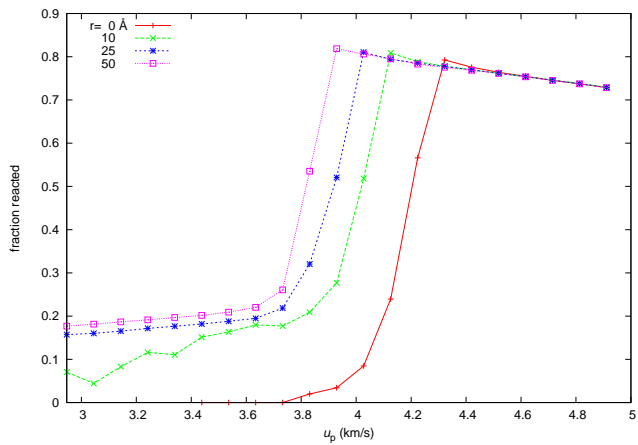}
\caption{Average maximum extent of reaction at each piston velocity and void radius.  Values in the plateau indicate detonation.}
\label{rextent}
\end{figure}

Each of 79 radius-velocity pairs was simulated 20 times, each until the shock broke out at the free surface.
In each simulation, the maximum number of reacted atoms observed (typically also the final count) was noted.  The average count for each \rup, as a fraction of the total number of atoms, is given in \fref{rextent}.  Even when the reaction consumes the entire sample, the conversion to $\mathrm A_2$ and $\mathrm B_2$ is not complete; reaction extents $>70\%$ indicate reaction of the entire sample (and thus detonation, since the reaction reached the shock front).  At velocities above 4.4~km/s, detonation is inevitable regardless of void size (or even presence); the reaction extent decreases slightly as velocity increases further because the equilibrium at the higher energy states is less completely reacted.

At velocities of 3.0--3.6~km/s, the hotspot consistently establishes a growing deflagration in the sample that merges with its periodic images and becomes planar but is left behind by the shock and does not become a detonation.  This process produces the consistent reaction extents of approximately 20\% that appear for all non-zero void sizes in \fref{rextent}.  The drop-off in the $r=10$~\AA\ data below $u_p=3.4$~km/s is largely due to failure to create a reacting hotspot, rather than due to reacting hotspots quenching.
For larger voids, that drop-off moves off below the $u_p$ range of the plot, and the ``shoulder'' of deflagration widens.  The step up to detonation moves more slowly and is of limited utility in identifying a critical velocity because more simulation time (with a larger sample) might allow some of the deflagrations to become detonations.

\begin{figure}
\centering\includegraphics[width=\figwidth]{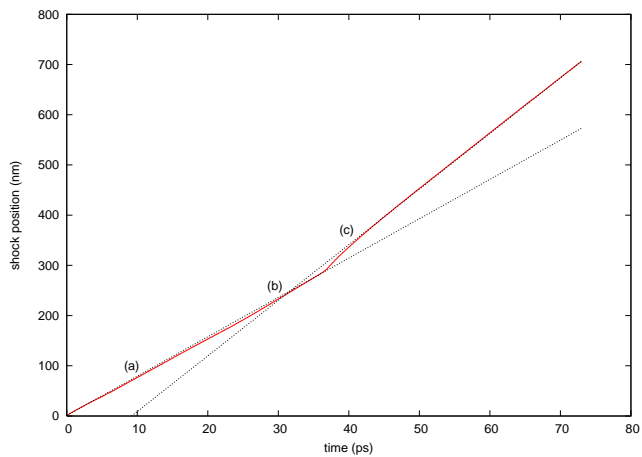}
\caption{History of the shock position in \fref{hvseq}'s simulation.  The times of those snapshots are marked.  The detonation transition is clearly visible; the dotted straight lines serve to illustrate the shock's acceleration before and after it.}
\label{shockex}
\end{figure}

The shock positions from a representative simulation that developed a detonation are shown in \fref{shockex} (the $\Delta z$-scale roughness is invisibly small).  Prior to the transition, reactions developing behind the shock accelerate it; this acceleration occurs even in the samples without voids, where one would expect the shock to accelerate only at transition (when the homogeneous initiation catches up from the piston face).  The conditions that inspire a detonation within such small homogeneous samples entail a chemical induction time so short that randomly distributed hotspots appear spontaneously and drive the detonation in the same fashion as in the heterogeneous case.

\begin{figure}
\centering\includegraphics[width=\figwidth]{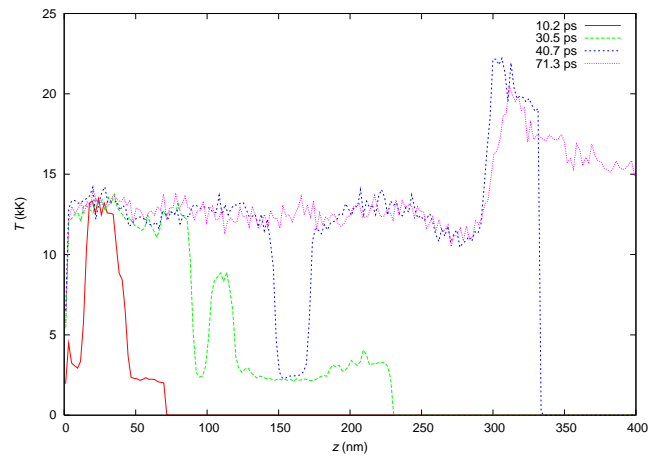}
\caption{Temperature profiles from \fref{hvseq}'s simulation over approximately the same region of the simulation.  The first three traces correspond to those snapshots; the last is from the end of the simulation and illustrates the stability of the features.  Temperatures of 2.5, 12.5, and 15~kK correspond to the inert shock, deflagration in the shocked region, and detonation respectively.}
\label{hvprof}
\end{figure}

Temperature profiles at four times during the same simulation are given in \fref{hvprof}.  The shock positions, the spread of several deflagrations throughout the shocked material, and the transition to detonation at $z=290$~nm are evident.  The persistent dip and spike astride the transition point seem to indicate that, when the reaction zone was just behind the shock but had not yet merged with it, the energy from the reaction went to accelerating the shock (and thus further heating the material being shocked) rather than to heating.  At the latest time, the boundary between the regions of deflagration and detonation remains well-defined, although the temperature spike has advected and diffused somewhat.

\begin{figure}
\centering\includegraphics[width=\figwidth]{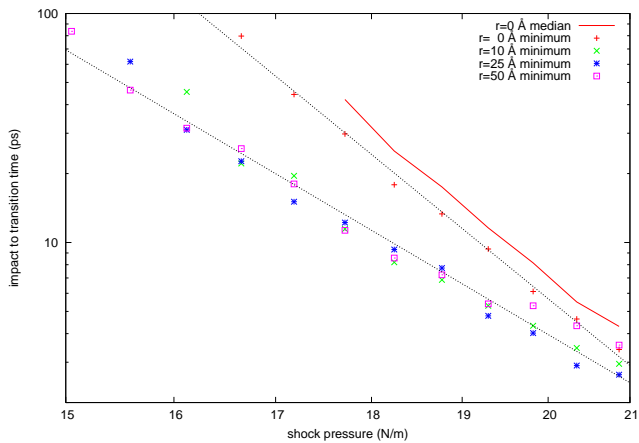}
\caption{\co Pop plot of the minimum times to detonation, with power-law fits to the no-void data and to all the data with voids.  The median times to detonation are also presented for the no-void case for illustration.}
\label{pop}
\end{figure}

The minimum transition times for each \rup\ (for which any detonations were observed) are given in \fref{pop}.  For comparison, the medians are also given for those voidless velocities for which a majority of simulations produced detonations; the medians for other void sizes behave similarly, with a nearly constant ratio between the minimum and median times.  We see pressure dependences of $p^{-13.77}$ with no void and $p^{-9.95}$ with any size of void, both much larger than the (space-pressure) exponents of $-1.6$ for PBX-9501\mcite{gibbs/popolato} and $-4.5$ for PBX-9502\mcite{info:lanl-repo/isi/000231885600019}\ecite.  This discrepancy may arise from the dimensionality of the system and the associated unusual pressure units.

At very high pressures, the void is irrelevant even to the promptness of the detonation, as many reactions are initiated directly upon the piston face.  (The largest voids even retard the high-pressure transitions, perhaps due to their greater distance from the piston.)   At lower pressures, the presence of a void greatly accelerates the development of a detonation by providing a guaranteed source of significant deflagration, but the size of the void seems not to significantly affect the subsequent positive feedback that develops the detonation.  However, the transition times for the three non-zero radii separate at low pressures, where the energy available from the hotspot becomes the determining factor in developing a detonation.  It happens that this change in the pressure exponent (for each radius) occurs just as the transition times become longer than the simulation, and so corresponds precisely to the lower limit of detonations in \fref{rextent}.

\section{Discussion}
We have obtained a simple form [\eqqref{cfit}, \fref{sigparam}] for the piston velocity needed to create a reacting hotspot from a void of a given radius in a two-dimensional ModelIV REBO high explosive crystal.  In the limit of large voids it appears sufficient to give each dimer just 70~meV of kinetic energy (7\% of the height of the repulsive core).  A difference of less than 400~m/s in piston velocity is expected to switch from 10\% to 90\% chance of ignition for voids with $r>10$~nm.

At higher velocities, we have observed a transition from steady deflagration to steady detonation in the range of $4.0\pm0.4$~km/s (an average kinetic energy of 1.1~eV per atom).  Its precise, radius-dependent location is non-trivial to establish, but these results suggest the form $u_c(r)\approx(3.8+0.34e^{-r/22\,\mAA})$~km/s.  As larger samples (and thus longer simulation times) might allow more detonations to finish developing, this expression is probably an overestimate.

The detonation initiation mechanism observed is neither the superdetonation associated with homogeneous explosives\ccite{beyond-standard} nor purely the ignition and growth of (the single rank of) hotspots, although such growth is observed (see \fref{hvseq}(a)).  Rather, the heat and pressure produced by the deflagration outpace it, encourage further reactions throughout the shocked material (see \fref{hvseq}(b)), and strengthen the shock until it ignites the material directly.  (The sample width [or distance between periodic images of the void] may affect the feedback; a larger sample provides more possible reaction sites but also more material over which to disperse the void's output.)  Extensions to this work may additionally consider the feedback resulting from the effects of a hotspot on further voids downstream. 
However, these results already suggest that even isolated nanoscopic features may directly affect bulk material behavior through the positive feedback inherent to energetic materials.

We have also observed the typical power-law dependence of detonation induction time on (two-dimensional) shock pressure, although the exponent seems to change at lower pressures.  To our knowledge, such a result has not previously been reported for an MD simulation.  It remains to be seen whether such a relationship holds in three-dimensional systems (whose pressures are more meaningful).

\ack
This report was prepared by Los Alamos National Security under contract no.~DE-AC52-06NA25396 with the U.S. Department of Energy.  Funding was provided by the Advanced Simulation and Computing (ASC) program, LANL~MDI contract~75782-001-09, and the Fannie and John Hertz Foundation.
\end{acknowledgments}

\end{document}